\documentstyle[aps,times]{revtex}
\begin{document}
\twocolumn[ {\bf Comment on ``The Transition Temperature of the Dilute
Interacting Bose Gas'' and on ``Transition Temperature of a Uniform Bose
Gas'' }\\
\smallskip
]
The transition temperature of a uniform, weakly interacting Bose gas has
received considerable interest in the past months, with at least three
contributions appearing in these pages alone.  In \cite{Baymetal} Baym
{\it et al.} argued that because of infrared divergences, the shift in
temperature due to the weak repulsive interaction relative to a free
Bose gas cannot be obtained in perturbation theory, no matter how weak
the interaction.  To regularize the infrared divergences, they derived a
self-consistent equation for the energy spectrum of the elementary
excitations and estimated the temperature shift as 
\begin{equation} 
\frac{T_{\rm c} - T_0 }{T_0} = c_0 \left(n a^3 \right)^{1/3},
\end{equation} 
with $c_0 \approx 2.9$.  Here, $T_0 = (2 \pi \hbar^2/m k_{\rm
B})[n/\zeta(3/2)]^{2/3}$, with $\zeta$ the Riemann zeta function, is the
transition temperature of an ideal Bose gas with particle number density
$n$, and $a$ is the s-wave scattering length characterizing the
repulsive interaction between the particles.

In \cite{Huang_PRL} Huang addressed the same issue using a virial
expansion of the equation of state.  He arrived at the conclusion that
the shift in $T_{\rm c}$ increases with the square root of the
scattering length, rather than linearly with $a$.

In this Comment we wish to point out that first, contrary to the claim
by Baym {\it et al.} \cite{Baymetal}, the shift in temperature can be
calculated perturbatively, as was done in \cite{effbos}.  And second,
that the regime considered by Huang \cite{Huang_PRL} does not describe
the Bose-Einstein condensed phase.

Specifically, he confines himself to the regime where the fugacity $z =
\exp(\mu/k_{\rm B}T)$ is smaller than unity, corresponding to negative
values of the chemical potential $\mu$.  This is because the used
expression for the pressure as function of $z$ becomes complex for
$z>1$.  However, the broken-symmetry phase of a weakly interacting Bose
gas is, unlike He-II, characterized by a {\it positive} value of $\mu$.
In particular, at $T=T_{\rm c}$ perturbation theory gives \cite{Popov}
$\mu = 8 \pi \hbar^2 a n/m$ to linear order in $a$.  Note that in a free
Bose gas $\mu$ is either negative (in the gas phase) or zero (in the
condensed phase).

In \cite{effbos} the shift in temperature, among other things, was
obtained perturbatively by calculating the pressure to the one-loop
order in a grand canonical ensemble as in \cite{Popov}, and expanding it
in a high-temperature series.  The justification of the high-temperature
expansion was given {\it a posteriori} by the observation that the
leading term in the expression for the transition temperature is of
order $(\mu/a)^{2/3}$, where we recall that $\mu$ is an independent
variable in a grand canonical ensemble.  Since this is large for $a$
small, the high-temperature expansion is consistent with the
weak-coupling assumption of perturbation theory.

The emerging infrared divergences were regularized by analytic
continuation, frequently used to regularize ultraviolet divergences.
Finally, using the equation of state to swap the chemical potential for
the particle number density as independent variable---which is more
appropriate from an experimental view point---we found that the shift in
temperature is given by Eq.\ (1), with
\begin{equation} \label{per}
c_0 = - \frac{8}{3} \frac{\zeta(\frac{1}{2})}{\zeta^{1/3}(\frac{3}{2})}
\approx 2.83.
\end{equation}
The estimate of Baym {\it et al.} is remarkably close to this
perturbative result.

Earlier path-integral Monte Carlo simulations \cite{GCL} restricted to
small (up to 216) particle numbers gave a value $c_0 = 0.34 \pm 0.06$
almost an order of magnitude smaller that the perturbative result
(\ref{per}).  More recent simulations \cite{HK} on larger systems
containing of the order of $10^4$ particles, gave a value $c_0 = 2.3 \pm
0.25$ in reasonable agreement with Eq.\ (\ref{per}).

In closing, we remark that in \cite{effbos} also a similar
high-temperature expansion and regularization of infrared divergences as
discussed here was applied to the BCS theory of superconductivity, and
shown to reproduce, among other things, the known result for the BCS
transition temperature.

I thank G. Baym, K. Huang and F. Lalo\"e for useful correspondence, and
M. Krusius for helpful advice.  This work was funded in part by the EU
sponsored programme Transfer and Mobility of Researchers under contract
No.\ ERBFMGECT980122.

\bigskip
\noindent
{\small Adriaan M. J. Schakel\\ Low Temperature Laboratory \\ Helsinki
University of Technology \\ P.O. Box 2200, FIN-02015 HUT, Finland \\  and
\\ Institut f\"ur Theoretische Physik \\ Freie Universit\"at Berlin \\
Arnimallee 14, 14195 Berlin, Germany \\ e-mail:
schakel@physik.fu-berlin.de \\

\noindent
PACS numbers: 03.75.Fi, 05.30.Jp}

\end{document}